October 2020

# Enhancing the role of academic librarians in conducting scoping reviews

Peter Kokol
peter.kokol@um.si

Jernej Zavrsnik dr
*Dr Adolf Drolc Healthcare Centre*, jernej.zavrsnik@zd-mb.si

Marko Turcin
*Dr Adolf Drolc Healthcare Centre*, marko.turcin@zd-mb.si

Helena Vošner Blažun
*Dr Adolf Drolc Healthcare Centre*, helena.blazun@zd-mb.si

# Enhancing the role of academic librarians in conducting scoping reviews

Peter Kokol[1], Jernej Završnik[2], Marko Turčin[2] and Helena Blažun[3]

[1] University of Maribor, Faculty of Electrical Engineering and Computer Science, Koroška 46, 2000 Maribor, Slovenia, peter.kokol@um.si

[2] Dr Adolf Drolc Healthcare Centre, Ulica talcev 9, 2000 Maribor, Slovenia,

[3] Community Healthcare Centre Dr. Adolf Drolc Maribor, Maribor, Slovenia
Faculty of Health and Social Sciences Slovenj Gradec, Slovenj Gradec, Slovenia
Alma Mater Europaea, Maribor, Slovenia

Corresponding author: Peter Kokol, PhD, MD; University of Maribor, Faculty of Electrical Engineering and Computer Science, Koroška cesta 46, 2000 Maribor, Slovenia, peter.kokol@um.si; +386 2 220 7457, ORCID: 0000-0003-4073-6488

**Abstract**

Information exposing, in conjunction with technological innovations and the emergence of social media, altered the traditional roles of academic libraries and enabled librarians to become necessary partners in research. The role of academic librarians in conducting systematic reviews is well recognised, however, their role in conducting scoping reviews is not yet well established. Nevertheless, we propose that, in more and more frequent situations when it is not feasible to read and analyse all relevant literature to be scoped manually, librarians employ bibliometric analysis and mapping to visualise and chart literature content. Our study demonstrated that science landscapes induced automatically by bibliometric mapping software could serve as a tool to visualise and chart the content of relevant literature when conducting the fourth step of scoping reviews. Additionaly science landscapes can help also serve to help improve the decision strategies when conducting scoping reviews.

**Introduction**

Scientific knowledge doubling (Bornmann, Anegón, & Leydesdorff, 2010; McDeavitt, 2014; R. Buckminster Fuller, 1983), in conjunction with information explosion, technological innovation, ease of access to information, new methods of scholarly communication and use of social media, altered the traditional roles of academic libraries and librarians (Loesch, 2017; Raju, 2014).

The role of academic librarians in conducting systematic reviews is well recognised (Nicholson, McCrillis, & Williams, 2017; Spencer & Eldredge, 2018). However, establishing their role in conducting scoping reviews is a new challenge. Scoping review (Pham et al., 2014) is an increasingly popular method to synthesise literature and knowledge. A literature search in Scopus (Elsevier, Netherlands) showed that the number of scoping reviews increased almost six fold in the last five years – from 137 occurrences in 2013 to 830 in 2017. Many librarians have noticed this trend, however, they also encountered some problems: (1) There is no universal definition of what scoping reviews are, and (2) While there is an agreed sequence of conducting scoping reviews, there is no established best way how to do these steps. While Morris, Boruff, & Gore (2016) defined tasks which can be performed by Health librarians in conducting scoping reviews, it is our belief that their role might become even more important, when manual reading and analysis of selected relevant literature might become too time consuming, or not economically feasible. We propose that, in such situations, librarians integrate the task defined by Moriss, Boruff and Gore with another of the new library services, namely bibliometric analysis (Åström & Hansson, 2013). In that manner, Health librarians can use bibliometrics as an additional tool to support researchers in analysing, visualising and charting the content of relevant publications. The purpose of this article is to demonstrate the feasibility of the above idea in scoping the recent research in the Paediatric Cardiology and Health Economics fields.

**Related literature**

*The new role of librarians*

Historically, the shift in services offered by academic libraries began in the late 1980s when Altmann (1988) argued that traditional division of public and technical services in an academic library was being challenged by technological changes and budget restrictions. Tyckoson (1991) proposed that libraries have to change from the ownership to an access model. Griffiths (1995), Melchionda (2007), Gruca (2010), Brindley (2010) and Sun et al (2011) claimed that librarians have to obtain skills and competencies for managing new ICT technologies, like the Internet, Web, and eLearning services in academic libraries.

In their new role, academic librarians can become necessary partners in research (Bedi & Walde, n.d.; Brahmi & Kaplan, 2017), consultants (Eddy & Solomon, 2017), data librarians (Koltay, 2017), add rigor to research (Sheffield, Refolo, Petanceska, & King, 2017), become embedded librarians (Cooper & Crum, 2013), support grant applications (Edmunds, Wright, & King, 2017), cooperate in conducting systematic and scoping reviews (Morris et al., 2016; Spencer & Eldredge, 2018), perform fact – checking (Lebeau, 2017), support evidence based practice in clinical education (Simons, Davidson, & Forrest, 2015), become repository managers (Cassella & Morando, 2012) and, finally, perform bibliometric studies (Åström & Hansson, 2013; González-Alcaide & Poveda-Pastor, 2018). In a recent study on 150 academic Health Science libraries listed in the AAHSL Membership Directory, Zhang (2018) showed that, among them, 82% provide citation management services, 70% writing guide and scholary publishing services, 55% copyright, fair use and author right service, 50% NIP Public Access Policy Compliance, 40% research data services, 40% research assessment services, 39% systematic review services and 35% grant writing and funding search services.

*Use of bibliometrics in medical research*

Bibliometric activities can be traced back to 1896, when Campbell (Campbell, 1896) published a book on the theory of national and international statistical Bibliography. In 1917, Cole and Eales (1917) performed the first counting of the number of publications per country in their analysis of international scientific activities in Comparative Anatomy and, thus, introduced the use of Bibliometrics to medicine. Lotkas paper on the frequency distribution of scientific productivity (Lotka, 1926), Gross paper on the role of a college library in chemical education (Gross & Gross, 1927) and Allens citation analysis of three periodicals for mathematicians (Allen, 1929) were the next historical roots of Bibliometrics. In 1934, Otlet (1934) introduced the term *bibliologie,* which could be translated to bibliometrics in his book about the dissemination of knowledge. The more intensive bibliometric research started with Garfield's (Garfield, 1955) seminal paper, in which he proposed the system how to eliminate the "uncritical citation of fraudulent, incomplete or obsolete data". Next, Price (de Solla Price, n.d.) proposed a tool to use science to analyse science and became the "father" of Scientometrics. In 1969, Pritchard (Pritchard, 1969), to disambiguate the term "statistical bibliography", defined Bibliometrics as the application of mathematics and statistical methods to books and other media. Since then, Bibliometrics has become widely used, especially in medicine. Indeed, Lewison and Devey (1999) described the role of Bibliometrics in medical research as "Bibliometrics is to scientific papers as Epidemiology is to patients." Bibliometrics became a medical subject heading in Medline in 1990. Since then, research literature production increased from 6 articles to 452 in 2015, when the production achieved its peak (Web of Science, Clarivate analytics).

*Bibliometric mapping*

In the scope of Bibliometrics, bibliometric mapping (Sinkovics, 2016) is used to analyse and visualise the structure, patterns and content of research literature in the form of bibliometric maps and science landscapes. Bibliometric mapping is often done using a popular mapping tool called Visualisation of Similarities (VOS) (vanEck & Waltman, 2010) (Leiden University, Netherlands) The advantage of VOSViewer is that the selection of terms on induced maps is based on proven empirical calculations, and that labelling and clustering of terms are performed automatically using dedicated algorithms (Hamilton et al., 2014). VOSviewer software merges terms that are closely associated into clusters, denoted by the same colour. On the other hand, proximity of terms can be interpreted as an indication of their relatedness. VOSViewer has been used successfully in several published studies related to Health (Kokol, Vošner, & Železnik, 2017; Peykari et al., 2015; Sweileh, 2017; Zhao et al., 2018).

*Scoping reviews*

Scoping reviews are becoming increasingly popular as a form of knowledge synthesis especially in health related research fields with emerging evidence (Levac, Colquhoun, & O'Brien, 2010). Scoping reviews can be described as a process of mapping the existing literature or evidence base (Arksey & O'Malley, 2005). Scoping and systematic reviews are similar in the manner that they both use rigorous and transparent methods to identify and analyse relevant literature to answer a research question (Pham et al., 2014). However they also differ considerably in several ways: The research question is narrower, study search criteria are defined at the outset, data extraction is very detailed and quantitative synthesis is done frequently (Armstrong, Hall, Doyle, & Waters, 2011). Additionally, scoping reviews deal with broader topics, might include different types of studies, do not evaluate the quality of the studies, create narratives to sum the results, and consider resources limitations like cost and time (Sarrami-Foroushani, Travaglia, Debono, Clay-Williams, & Braithwaite, 2015). Scoping reviews are performed when there are research gaps in specific areas of research, to disseminate the research results, to clarify a complex concept, as a preliminary step before conducting systematic reviews to explore the extent of literature, and, finally, when informative summary of a research in a specific professional area is needed (Arksey & O'Malley, 2005; Armstrong et al., 2011; Levac et al., 2010; Peterson, Pearce, Ferguson, & Langford, 2017; Tricco et al., 2016). Recently, scoping reviews are also employed for knowledge translation (Schultz et al., 2018). Arksey and O'Maley (2005) proposed a six step methodological framework to perform scoping reviews. This framework was later extended by Levac, Colquhoun & O'Brien (2010) and Joanna Briggs Institute (Peters et al., 2017).

Since 1990 when conducting systematic reviews has become more frequent, librarians have played different roles in their preparation, like planning, searching and setting search filters, removing duplicate titles, publication selection, teaching others how to perform systematic reviews, research question formulation, use of analytical tools and peer review (Spencer & Eldredge, 2018). Focusing on scoping reviews, librarians can cooperate in the following tasks (Morris et al., 2016):

- Support researchers in understanding the relation of the breadth of the research question in relation the number of retrieved items;
- Develop a search strategy to find the relevant studies;
- Select the studies that are relevant to the research question(s) and obtain hard to find publications;
- Help to chart, organise, summarise, report and export the results;
- Help to describe the methodology used.

**Methodology**

The analysis of related research literature revealed that new age academic libraries already provide/plan bibliometric services and support conducting scoping reviews. In our paper, we would like to show that, if needed, extending the current academic library bibliometric service with bibliometric mapping and integrating the enhanced service into the process of conducting a scoping review, can make this process more effective and the librarians` role even more useful.

To demonstrate the feasibility of the above idea two Proof Of Concept Studies (POCS) were performed. The first POCS was aimed to get an overview of the research in Cardiovascular Paediatrics, and the second to summarise the recent research in Health Economics. Both POCS were aimed to answer the question:

- Can VOSViewer induced cluster maps serve as a basis to chart, organise and summarise the content of relevant research literature in conducting scoping reviews?

*Search strategy and data extraction*

The papers were harvested from the PubMed. The search was performed on paper titles, abstracts and keywords, using the MeSH defined termsPa*pediatric Cardiology* for the first and *Health Economics for the* second case study. For Paediatric Cardiology, the search was limited to the period 2012 –2016 (the search was done in November 2017) and for the Health Economics study, to the period 2015 – 2017 (the search was done in July 2018).

*Analysis, visualization and charting the content*

We used VOSViewer to visualise and chart the content of relevant research literature in the form of a science landscape. Using a customised Thesaurus file, we excluded the common terms like study, significance, sample, baseline, group, experiment, trial, or country, and eliminated geographical names . Similarly, synonyms like cost, health cost or Paediatric and Paediatrics were fused into one term. The science landscapes were used to identify interesting associations and relations. The process was iterative, performed by an interdisciplinary team of Information, Health Science, Health Economy and Paediatrics Cardiology experts.

**Results**

*POCS 1: Paediatrics Cardiology*

The search resulted in 8,550 publications. During the first iteration, we induced the science landscape presented in Figure 1. To help the domain experts to generate the feedback, we identified the following associations and grouped them in the scope of five themes, based on the clusters induced by VOSViewer:

- Critical congenital diseases (blue colour)
    - Critical Congenital Heart Disease (CHD), associated with high cost, mortality, survival and safety;
    - AtrioVentricular Nodal Re-entry Tachycardia (AVNRT), associated with ExtraCorporeal Membrane Oxygenation (ECMO) and Risk Adjustment in Congenital Heart Surgery (RACHS);
    - Heart transplant, associated with CardioPulmonary Bypass (CPB), cardiac allograft vasculopathy and acute kidney injury;
    - Fontan patient, associated with Fontan circulation.
- Congenital Heart Diseases (red colour)
    - Patent Ductus Arteriosus (PDA), associated with Amplatzer Duct Occlude (ADO) closure, coil occlusion and stent implantation;
    - Ventricular septal defect, associated with high pulmonary vascular resistance , Anomalous Left Coronary Artery from the Pulmonary Artery (ALCAPA), TransThoracic Echocardiogram (TTE) and Tetralogy of Fallot ;
    - Pulmonary Arterial Hypertension (PAH), connected to Congenital Heart Disease associated with different syndromes.
- Prenatal heart diseases (yellow colour)
    - Congenital Heart Disease (CHD) in foetal age, associated with MRI and Body Surface Area (BSA);
    - Tissue Doppler imaging, associated with strain rate echocardiography and exercise ECHO.
- Acquired heart diseases (green colour)

- o Obesity, associated with Body Mass Index, ventricular mass, arterial stiffness, CIMT testing, left ventricular motility injury, diabetes, metabolic syndrome and atherosclerosis;
- o Dilated cardiomyopathy, associated with potential stress and left ventricular non-compaction cardiomyopathy;
- o Phenotype, associated with cell, cardiomyocyte and mutation;
- o Hypertension, associated with fibrosis and cardiac dysfunction;
- o *Coarctation of Aorta (CoA)*, associated with ToF and BAV.
- Long QT syndrome (violet colour), associated with seizure, beta blocker and electrocardiogram.

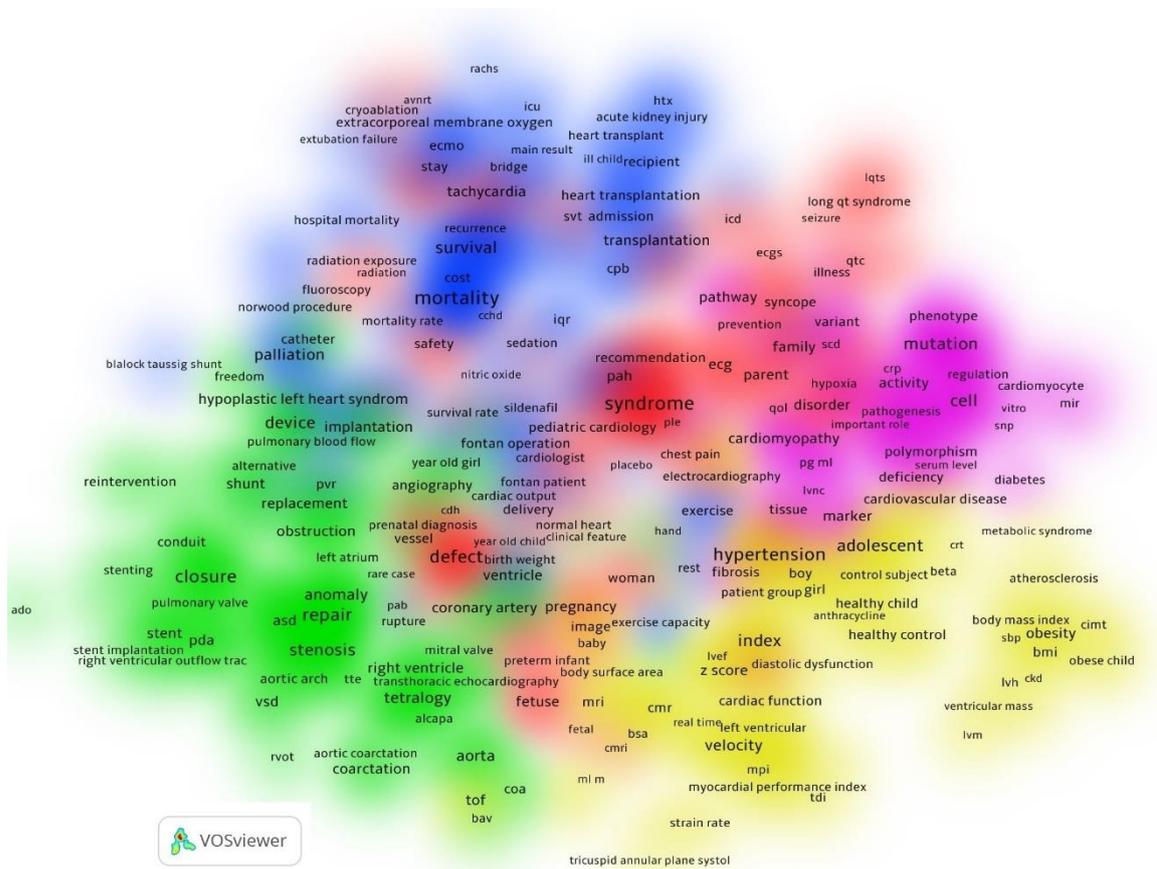

Figure 1. The science landscape of Paediatric Cardiology for the period 2012–2016

Domain experts reviewed the science landscape and the textual summary and gave us the following feedback:

- There are some doubtful associations: AVNRT and ECMO, VSD and ALCAPA, AVNRT and RACHS, Coarctation of the Aorta being associated with TOF and BAV
- There are terms which should be removed: Recommendation, control subject, patient group, defect, syndrome, boy, delivery, year old girl, children
- Synonyms should be merged: Cardiac surgery, surgery, and congenital heart surgery; neonate, infant and new-born
- Medical and technical inconsistencies should be corrected: Doppler echocardiography and TEE are not treatments, they are diagnostic modalities, Why list aortic coarctation and ASD as themes separately from CHD? How is ventricular mass a treatment?
- General remarks: The landscape is not easy to read, it should be presented in another format! The textual form of relations and associations is good for discussions, but not necessary in the "final chart".

After taking into account the observations from the feedback, we performed another iteration and induced the science landscape presented in Figure 2. This landscape was approved by domain experts as a suitable foundation to collate and summarise the main research activities and themes in Paediatric Cardiology research (Step 5 in conducting scoping reviews).

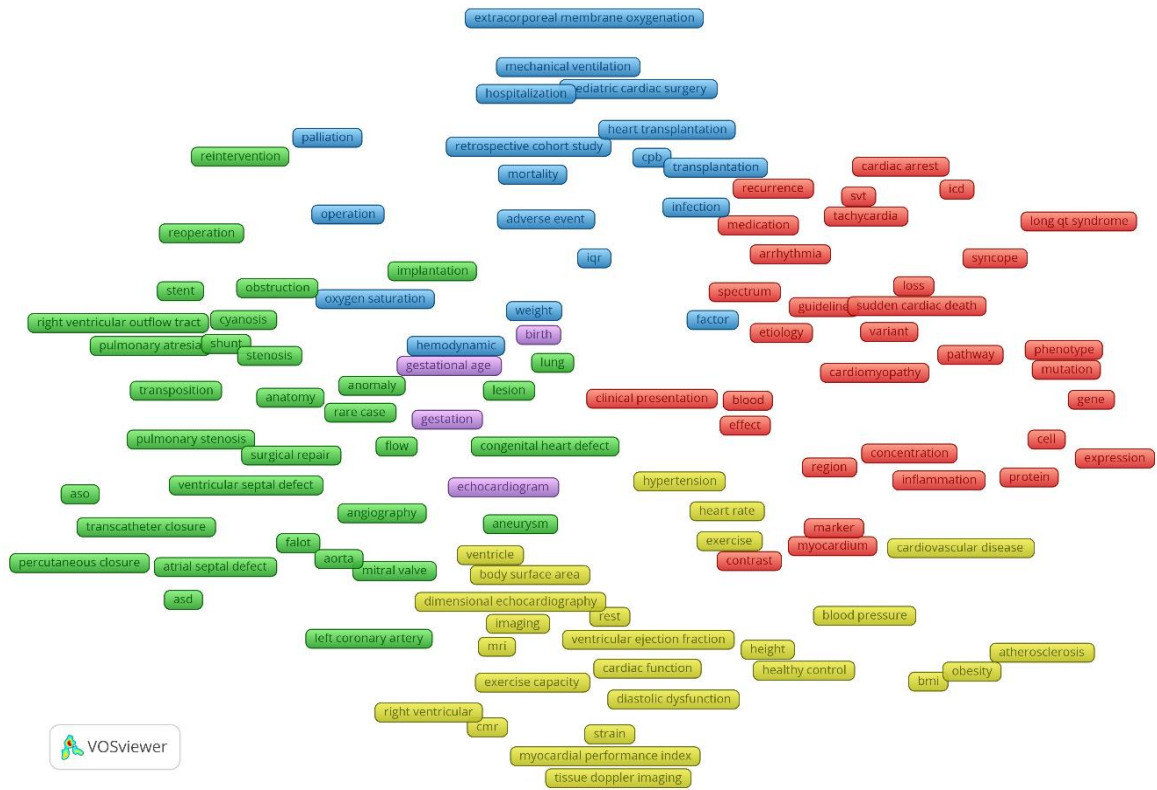

Figure 2. The science landscape of Paediatric Cardiology for the period 2012–2016 after reiterations

## POCS 2: Health Economics

The search resulted in 20,501 papers, and their analysis in the science landscape presented in Figure 3. Based on the experiences from the first study, we tried a modified approach in the second study. Instead of doing the thematic analysis in advance, we performed it in close cooperation with experts and named the themes together. The domain experts were happy with the derived themes, however, they requested that, based on the theme terms, we derive a list of search keywords for each theme for further scoping of individual themes. The themes and keywords are listed below:

- Decision-making based on cost analysis, discrete choice modelling and technology assessment in Oncology (red colour)
  *Keywords*: Cost analysis, Oncology, technology assessment, hospitalization, chemotherapy, palliative care, economic burden and reimbursement.

- Economic evaluation by different kinds of cost - effectiveness, cost - utility and cost - benefit analysis (green colour)
  *Keywords:* Quality Adjust Life years (QALF), cost analysis. Effectiveness, benefit, and prevention.
  *Excluding keyword*: Randomised control trial.

- Economic impact of risks of non-healthy life style (dark blue colour)
  *Keywords:* Economic impact, risk, non-health life style, education, Epidemiology and health service research.

- The cost of health related quality of life in integrated long term care (yellow and light blue colours)
  *Keywords:* Health Related Quality of Life (HRQL), long-term care, comorbidity, adherence,
  *Excluding keywords*: HIV, Hepatitis c and asthma

- Macro - economic aspects of health care (purple colour)
  Keywords: Health care, health insurance, health reform, macro economic aspect

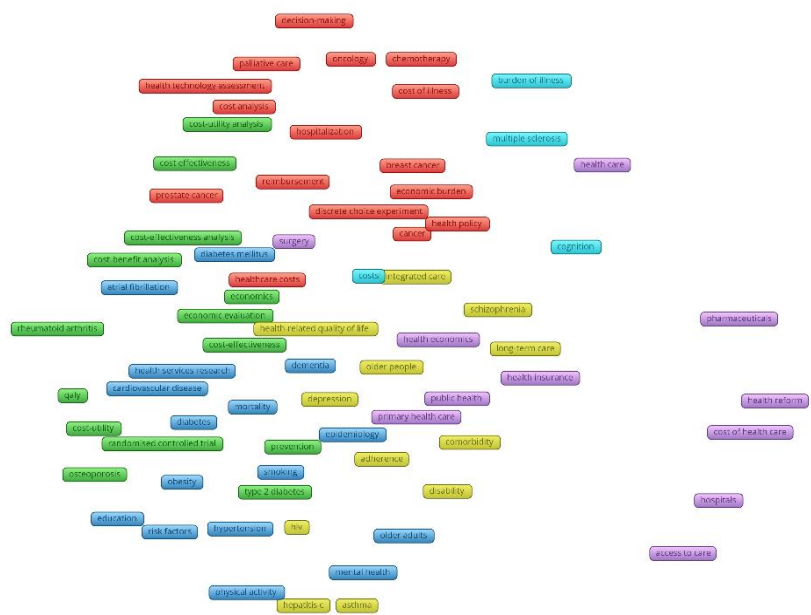
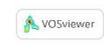

Figure 3. The science landscape of Health Economics research for the period of 2013 - 2017

**Discussion and Conclusion**

Our study demonstrated that scientific cluster landscapes induced by VOSViewer could serve as a tool to visualise and chart the content of relevant literature when conducting scoping reviews. It also showed that cluster landscapes can serve as a tool to elaborate the search strategies in conducting scoping reviews. Some academic libraries already perform bibliometric analyses (Åström & Hansson, 2013; González-Alcaide & Poveda-Pastor, 2018), and some academic librarians already support conducting scoping reviews (Morris et al., 2016). Consequently, academic librarians already possess or will possess bibliometric analysis and conducting scoping review skills, hence establishing an academic library service which will perform bibliometric mapping based support in conducting scoping reviews is a viable option. Additionally, VOSViewer is an open licenses software package and can, thus, be used in academic libraries without additional costs. For a librarian with basic bibliometric analysis skills it is easy to learn and use.

vanEck, N. J., & Waltman, L. (2010). Software survey: VOSviewer, a computer program for bibliometric mapping. *Scientometrics*, *84*(2), 523–538.

Zhang, Y. (2018). Provision of Research Assessment Services in Academic Health Sciences Libraries. https://doi.org/10.7282/T3SJ1PZZ

Zhao, J., Yu, G., Cai, M., Lei, X., Yang, Y., Wang, Q., & Zhai, X. (2018). Bibliometric analysis of global scientific activity on umbilical cord mesenchymal stem cells: A swiftly expanding and shifting focus. *Stem Cell Research and Therapy*, *9*(1). https://doi.org/10.1186/s13287-018-0785-5
17